\documentclass{iopart}
\usepackage{iopams}
\usepackage{bbm}
\usepackage{dsfont}
\usepackage{amsthm}
\usepackage{amscd}
\usepackage[ansinew]{inputenc}
\usepackage{cite}
\usepackage[dvips]{graphicx}

\newcommand{\eqref}[1]{(\ref{#1})}
\newcommand{\nn}{\nonumber}
\newcommand{\be}{\begin{equation}}
\newcommand{\ee}{\end{equation}}
\newcommand{\ba}{\begin{eqnarray}}
\newcommand{\ea}{\end{eqnarray}}

\newcommand{\rf}[1]{(\ref{#1})}

\newcommand{\J}{v}
\newcommand{\U}{c}
\newcommand{\g}{g}
\newcommand{\s}{s}
\newcommand{\half}{\frac{1}{2}}
\newcommand{\cP}{\mathcal{P}}
\newcommand{\cT}{\mathcal{T}}
\newcommand{\cH}{\mathcal{H}}
\newcommand{\cE}{\mathcal{E}}
\newcommand{\om}{\widetilde\omega}

\def\p{\partial}
\newcommand{\gcs}{|\alpha\rangle}
\newcommand{\hcs}{\langle\alpha |}
\newcommand{\symp}{\mathit\Omega}

\begin{document}
\title[Classical limit of non-Hermitian
quantum dynamics]{Classical limit of non-Hermitian
quantum dynamics --  a generalised canonical structure}
\author{Eva-Maria Graefe$^{1,2}$, Michael H\"oning$^1$ and Hans J\"urgen Korsch$^1$}
\address{${}^1$ FB Physik, TU Kaiserslautern, D--67653 Kaiserslautern, Germany\\
${}^2$ School of Mathematics, University of Bristol, Bristol BS8 1TW, United Kingdom}
\ead{e.graefe@bristol.ac.uk, hoening@rhrk.uni-kl.de, korsch@physik.uni-kl.de}
\begin{abstract}
We investigate the classical limit of non-Hermitian quantum dynamics arising 
from a coherent state approximation, and show that the resulting 
classical phase space dynamics can be described by generalised ``canonical'' 
equations of motion, for both conservative and dissipative motion. 
The dynamical equations combine a symplectic flow associated 
with the Hermitian part of the Hamiltonian with a 
metric gradient flow associated with the anti-Hermitian 
part of the Hamiltonian. We derive this structure of the 
classical limit of quantum systems in the case of a Euclidean phase space geometry. 
As an example we show that the classical dynamics of a damped and driven oscillator 
can be linked to a non-Hermitian quantum system, and investigate the 
quantum classical correspondence. Furthermore, we present an example 
of an angular momentum system whose classical phase space is spherical and 
show that the generalised canonical structure persists for this nontrivial phase space geometry.     
\end{abstract}

\submitto{\JPA}

\section{Introduction}
While quantum mechanics traditionally demands the 
Hamiltonian be Hermitian for the description 
of closed systems, non-Hermitian Hamiltonians have proven 
useful for the description of open systems involving decay, 
scattering, and transport phenomena \cite{Gamo28,Datt90b,Okol03,Berr04,Nare03,Mois98}. 
Here one uses Hamiltonians whose eigenvalues have 
negative imaginary parts which leads to a monotonic 
decrease of the overall probability in time. Although in most cases non-Hermitian 
Hamiltonians are introduced heuristically, they can be incorporated in 
a mathematically satisfactory way starting from the system coupled 
to a continuum of states (see \cite{Datt90b,Okol03,Maha69} and references 
cited therein). Furthermore,  non-Hermitian $\cP\cT$ symmetric 
Hamiltonians can have a purely real spectrum in certain parameter 
regions and can be used to define a consistent description of 
closed quantum systems \cite{Bend98,Bend02b,Bend07}. Non-Hermitian 
quantum dynamics is sometimes regarded as a mere perturbation 
to Hermitian dynamics, adding an overall decay. However, 
non-Hermitian dynamics is in general very different from the familiar unitary time evolution. 
Much of this difference is due to the non-orthogonality of the non-Hermitian eigenfunctions, 
which can even degenerate at ``exceptional points'' in parameter space. The 
various intriguing effects associated with non-Hermitian quantum evolution attracted considerable 
attention recently, both from theoretical and from experimental side 
\cite{Ober96,Berr98,Muss08,Wier08,Toma07,Toma08,Hata08,Klai08,Elga07}.
The field is still in an early stage of its development and many generic features are 
far from being fully understood. 

Little is known about the quantum-classical correspondence 
of systems that are open in the above sense. Considerable insights have 
recently been gained from the study of toy models 
described by subunitary or truncated 
unitary operators associated with open quantum maps, 
where a fractal Weyl law has been found \cite{Lu03,Nonn08,Keat06}. In the 
general case, however, even the classical analogue of 
non-Hermitian quantum theories has 
hitherto remained an open question. Complex extensions 
of classical dynamics in terms of complex coordinates 
and momenta have been studied 
in the contexts of both Hermitian and non-Hermitian 
theories \cite{Xavi96,00complex,Bend01,Bend07b}. 
But these systems cannot be regarded as representing 
the classical limits of non-Hermitian quantum systems for 
which momentum and coordinate operators 
have real expectation values. On the other hand, 
in attempts to quantising classical damped motion, 
a possible link to non-Hermitian Hamiltonians has been 
observed  \cite{Dekk75,Raje07,Geic89}, but studies in this 
direction have not been substantiated, 
there is no generic structure available, and none of the 
approaches has been able to provide the desired quantum-classical 
correspondence in a rigorous sense. Recently a generalised classical 
approximation using algebraic coherent states \cite{Yaff82,Pere86,Zhan90} 
for a non-Hermitian Bose-Hubbard dimer \cite{Hill06,08PT} was 
introduced in \cite{08nhbh_s} in the context of Bose-Einstein condensates. 

Here we show that there is a generalised canonical structure 
that can be derived as the classical limit of non-Hermitian quantum 
dynamics. This structure incorporates a metric gradient flow generated by the 
anti-Hermitian part of the Hamiltonian. It is closely related 
to the canonical formulation of classical damped motion 
suggested in the past \cite{Kauf84,Morr86,Holm07}, as well as 
to the gradient flow appearing in thermal and statistical physics and 
control theory \cite{Bloc92,Brod96}. It should be emphasised 
that we do not start from a classical damped 
system and quantise it. We rather begin with the non-Hermitian 
quantum system and perform a coherent state approximation 
that is often used to define the classical limit in the Hermitian case. 
The generalised canonical structure is derived for arbitrary quantum 
systems having a Euclidean phase space. The Bose-Hubbard 
system studied in \cite{08nhbh_s}, however, can be regarded as an 
angular momentum system and thus the classical phase space is given by the 
Bloch sphere. Here we show that the corresponding classical dynamics 
can nevertheless be formulated according to the proposed generalised 
Hamiltonian structure. This gives some evidence that the proposed structure persists 
for more general phase space geometries.

The paper is organised as follows: In section \ref{s-nhqd} we give a brief introduction to 
non-Hermitian quantum dynamics, before we introduce the classical limit 
using coherent states in section \ref{sec-clas-lim}. We establish the 
generalised canonical structure for arbitrary 
one-dimensional quantum systems on a flat phase space, 
using Glauber coherent states. This derivation is analogous for higher dimensions. 
In section \ref{sec-damped-osc} we present a case study for a damped oscillator. 
In section \ref{sec_ang}, we briefly review the classical 
approximation \cite{08nhbh_s} for the angular momentum system studied 
in \cite{Hill06,08PT} and show that the generalised canonical structure 
can be used to describe the classical dynamics on the spherical 
phase space. We end with a summary and short outlook. 
\section{Non-Hermitian quantum dynamics}
\label{s-nhqd}
For a general quantum system with Hamiltonian $\hat
\cH$, not necessarily Hermitian, the dynamics of a pure
state $|\Psi\rangle$ is governed by the Schr\"odinger equation
\begin{eqnarray}
\rmi\hbar|\dot{\Psi}\rangle=\hat \cH|\Psi\rangle\,.
\label{schreq1}
\end{eqnarray}
We can decompose the Hamiltonian into a Hermitian and an
anti-Hermitian part via $\hat{\cH}=\hat H -\rmi \hat\Gamma$, with
$\hat H=\hat H^\dagger$ and $\hat\Gamma=\hat \Gamma^\dagger$, where
the damping term $\hat \Gamma$ is assumed to be non-negative. The special 
choice of the anti-Hermitian part $\hat \Gamma$ depends on the 
dynamics to be modelled, and the physical interpretation as a damping 
will become clear later. Under time evolution all states decay towards the 
subspace spanned by the eigenvectors of $\hat{\cH}$ with 
the smallest decay rate (imaginary part of the eigenvalue). 
Assuming a time independent Hamiltonian with a discrete 
spectrum, the time evolution can be expressed as
\be
|\psi(t)\rangle=\sum_n c_n\rme^{-\rmi E_nt/\hbar} |\varphi_n\rangle.
\label{psit-n}
\ee
Here $\hat{\cH}|\varphi_n\rangle = \cE_n|\varphi_n\rangle$, $\cE_n=E_n-\rmi \Gamma_n$
with real $E_n\le E_{n+1}$ and non-negative $\Gamma_n$. The expansion coefficients
$c_n$ can be obtained from the initial state by projection onto the
left eigenstates, i.e.~the eigenstates of $\hat\cH^\dagger$. If there are no 
real eigenvalues, every initial state approaches the most stable initially populated 
eigenstate in the long time limit:
\be
|\psi(t)\rangle
\longrightarrow c_0 \rme^{-\Gamma_0t/\hbar}\,\rme^{-\rmi E_0 t/\hbar} |\varphi_0\rangle\,,
\ee
where we have assumed that this is the (non-degenerate) ground state, and the norm approaches
\be
n(t)=\langle \psi(t)|\psi (t)\rangle \longrightarrow 
|c_0|^2 \rme^{-2\Gamma_0t/\hbar}\langle \varphi_0|\varphi_0\rangle\,.
\label{Ninf}
\ee

From the non-unitary state evolution (\ref{schreq1}) one
can immediately deduce a generalised Heisenberg equation of motion
\cite{Datt90b} for the diagonal matrix element of
an operator $\hat A$ (for simplicity we consider only the case
$\partial \hat A/\partial t=0$):
\begin{eqnarray}
\nn \rmi \hbar \frac{\rmd \,}{\rmd\, t}\langle \psi|\hat A|\psi
\rangle &=&\langle \psi|\hat A\hat{\cH}-\hat{\cH}^\dagger\hat
A|\psi \rangle\\ &=&\langle \psi|\,[\hat A,\hat H]|\psi \rangle -
\rmi\langle \psi|\,[\hat A,\hat \Gamma]_{\scriptscriptstyle +}|\psi
\rangle, \label{dattoli1}
\end{eqnarray}
where  $[\ ,\ ]_{\scriptscriptstyle +}$ is the
anti-commutator. Taking into account the fact that the norm is not conserved 
but decays according to
\begin{eqnarray}
\hbar\frac{\rmd \,n}{\rmd\, t}
=-2\langle \psi|\hat \Gamma|\psi \rangle \,,
\label{norm}
\end{eqnarray}
the dynamical equation of motion for the expectation value
of an operator $\langle \hat A\rangle =\langle \psi|\hat A|\psi
\rangle/\langle \psi|\psi \rangle$ reads
\begin{eqnarray}\label{GHE}
\rmi \hbar \frac{\rmd \,}{\rmd\, t}\langle \hat A \rangle
=\langle[\hat A,\hat H]\rangle - 2\rmi \,\Delta^2_{A\Gamma} \,,
\end{eqnarray}
with the covariance
$\Delta^2_{AB}=\langle {\textstyle \frac12}[ \hat A, \hat
B]_{\scriptscriptstyle +} \rangle
- \langle  \hat A \rangle \langle \hat B \rangle\,.$
In particular, for $\hat A=\hat H$ we have
\begin{eqnarray}
\hbar \frac{\rmd \,}{\rmd\, t}\langle \hat H \rangle
=- 2\,\Delta^2_{H\Gamma} \,.
\end{eqnarray}
This evolution equation simplifies in the special case
$\hat \Gamma = k\hat H$ with a non-negative real constant $k$ to
\begin{eqnarray}\label{GHE-H}
\hbar \frac{\rmd \,}{\rmd\, t}\langle \hat H \rangle
=- 2\,\big(\langle \hat H^2\rangle - \langle \hat H\rangle^2\big)\,.
\end{eqnarray}
This equation has been suggested by 
Gisin \cite{Gisi81a,Gisi81b}, however, with no connection to the
non-Hermitian Schr\"odinger equation (\ref{schreq1}). 
Instead, it was derived from the nonlinear quantum evolution equation 
\begin{eqnarray}
\rmi\hbar|\dot{\phi}\rangle
=\hat H|\phi\rangle
-\rmi \big( \langle \hat \Gamma\rangle -  \hat \Gamma\big)|\phi\rangle
= \hat H|\phi\rangle-\rmi  \big[|\phi\rangle\langle\phi|,\hat
\Gamma\big]|\phi\rangle
\label{schreq2}
\end{eqnarray}
with $\langle \hat A\rangle =\langle\phi |\hat A|\phi\rangle$ 
\cite{Gisi81a,Gisi81b,Gisi82,Gisi83a,Gisi83b}.
One can easily see that there is a direct relation between
the nonlinear dissipative quantum evolution equation (\ref{schreq2}) 
and the non-Hermitian Schr\"odinger equation (\ref{schreq1}).
A transformation to a renormalised state vector 
$|\phi\rangle=|\psi\rangle/\sqrt{n}$ using (\ref{schreq1}) and
(\ref{norm}), i.e.
$\hbar \dot n =-2\langle \hat \Gamma\rangle n$,
immediately leads to \rf{schreq2}.
Note that these nonlinear evolution equations can also be expressed in terms
of the density operator (see, e.g., \cite{Gisi81a,Gisi81b}), where they 
take the form of a double bracket flow frequently investigated 
in control theory and thermal physics \cite{Broc91,Bloc96,Brod08}. 
Furthermore, they can be extended 
to mixed states evolving towards thermal equilibrium
\cite{87dqs,92dqs,Bere01,Bere06}. Here, however, we intend to 
investigate the classical limit of the
non-Hermitian evolution (for a pure state), 
a topic only briefly addressed in \cite{Gisi81a}.

\section{A generalised classical limit for one-dimensional quantum  dynamics}
\label{sec-clas-lim} 
Often the classical counterpart of a quantum
system is obtained in a rather sloppy manner by simply ``taking the
hats off the operators'' and identifying the resulting quantities
with the classical ones. For a one-dimensional quantum system
describing a particle of mass $m$ with momentum $\hat p$ and position
$\hat q$ one can conveniently express the Hamiltonian in terms of harmonic oscillator
ladder operators 
\begin{eqnarray}
\hat{a}=\frac{m\omega\hat{q}+\rmi\hat{p}}{\sqrt{2m\hbar\omega}}\ ,\qquad
\hat{a}^\dagger= \frac{m\omega\hat{q}-\rmi\hat{p}}{\sqrt{2m\hbar\omega}}.
\end{eqnarray}
Coherent states $\gcs$ are defined by
\begin{equation}
\hat a\gcs=\alpha \gcs \,,\qquad \alpha =\frac{m\omega q+\rmi p}{\sqrt{2m\hbar\omega}}
\end{equation}
with $q=\hcs\hat q\gcs$ and $p=\hcs\hat p\gcs$, and the minimal 
uncertainty product $\Delta q \Delta q=\hbar/2$, 
which can be associated with a phase space point in the limit
$\hbar\to 0$. 

Applying the identification
\begin{eqnarray}\label{eqn-classical-approx1}
\sqrt{\hbar}\hat a \to z=\frac{m\omega q+\rmi p}{\sqrt{2m\omega}}\quad{\rm and}\quad
\sqrt{\hbar}\hat a^\dagger \to z^*=\frac{m\omega q-\rmi p}{\sqrt{2m\omega}},
\end{eqnarray}
replacing all operators in the Hamiltonian with their associated
c-numbers, and then identifying the result with a Hamiltonian
function on the classical phase space, one 
obtains the classical counterpart in this formulation.
The Heisenberg equations of motion for an operator $\hat F$ 
are implicitly replaced with the canonical equation of
motion in terms of Poisson brackets for the associated function
$F(q,p)$ of the classical canonical variables:
\begin{eqnarray}
\dot{\hat{F}}=\frac{1}{\rmi\hbar}[\hat{F},\hat{H}]\quad \to\quad
\dot F=\frac{1}{\rmi}\lbrace F,H\rbrace_{z,z^*}=\lbrace F,H\rbrace_{q,p},
\end{eqnarray}
with Poisson brackets $\lbrace \,,\, \rbrace$. Here we made use of the
complex formulation of classical mechanics in terms of $z$ 
and $z^*$ \cite{Stroc66}. This classical approximation can 
be motivated in the following way: 
For an operator function $\hat F$ the relations
\begin{equation}
[\hat{a}, \hat{F}]=\frac{\partial \hat{F}}{\partial \hat{a}^\dagger}\quad
{\rm and}\quad
[\hat{a}^\dagger, \hat{F}]=-\frac{\partial \hat{F}}{\partial \hat{a}}
\end{equation}
hold \cite{Loui90}. Therefore, we can express the
Heisenberg equations of motion for the ladder operators in the form
\begin{eqnarray}
\rmi\hbar\dot{\hat{a}}=\frac{\partial\hat{ H}}{\partial\hat{a}^\dagger}\quad
{\rm and}\quad
\rmi\hbar\dot{\hat{a}}^\dagger=-\frac{\partial\hat{H}}{\partial\hat{a}}.
\end{eqnarray}
We can than think of the identification \eqref{eqn-classical-approx1} 
as replacing the quantum operators by their expectation values in 
coherent states, for which expectation values can be factorised. 
This immediately yields the canonical equations of motion for $z$ 
and $z^*$, and the corresponding equations of motion for a 
function of the canonical variables in terms of the Poisson brackets. 
Thus, the classical approximation can be rephrased as the
assumption that an initially coherent state stays coherent
throughout the time evolution. This is analogous to the spirit of
the so-called frozen Gaussian approximation \cite{Hell81, Kluk86}
for the dynamics of a quantum system. Note that the 
approximation becomes exact if the Hamiltonian is 
linear in the group operators $\hat a^\dagger \hat a$, $\hat a$ and $\hat a^\dagger$ which
can be verified by explicit calculation of the action of 
the resulting time evolution operator on an initially coherent state. 
In a semiclassical context in addition one often symmetrises the appearing ladder 
operators to account for the zero point energy. Here, however, 
we deal with the \textit{classical} approximation for $\hbar\to0$ where 
the zero point energy vanishes identical and thus, we neglect issues of operator ordering. 
Further, while an additional constant changes the energy values, it leaves the 
dynamics invariant.

Let us now derive the general structure of the classical equations of
motion for a non-Hermitian Hamiltonian. 
For this we have to translate the equations of motion for the
expectation values of the ladder operators for an arbitrary
Hamiltonian to their classical counterparts using the coherent
state approximation.
We can express the anti-commutators appearing in the equation of
motion, making use of the identity \cite{Loui90}
\begin{eqnarray}\label{eqn-anticomm_a_F}
[\hat{a}, \hat{F}]_{+}=2\hat{F}\hat{a}+ \frac{\partial \hat{F}}{\partial \hat{a}^\dagger},\quad
[\hat{a}^\dagger, \hat{F}]_{+}=2\hat{a}^\dagger\hat{F}+ \frac{\partial \hat{F}}{\partial \hat{a}}
\end{eqnarray}
for an operator function $\hat F=\hat F(\hat a,\hat
a^\dagger)$. Inserting expressions (\ref{eqn-anticomm_a_F}) in the generalised Heisenberg equations
of motion \rf{GHE} we find
\begin{eqnarray}
\begin{array}{ll}
 \rmi\hbar \frac{d}{dt}\langle{\hat{a}}\rangle&=\phantom{-}\langle
 \frac{\partial\hat{H}}{\partial\hat{a}^\dagger}\rangle
-\rmi\big(\langle2\hat{\Gamma}\hat{a}+\frac{\partial\hat{\Gamma}}{\partial\hat{a}^\dagger}\rangle-2\langle\hat{a}\rangle\langle\hat{\Gamma}\rangle\big)\\
\vspace{-0.3cm}\\
 \rmi\hbar\frac{d}{dt}\langle{\hat{a}^\dagger}\rangle&=-\langle\frac{\partial\hat{H}}{\partial\hat{a}}\rangle-\rmi\big(\langle2\hat{a}^\dagger\hat{\Gamma}+\frac{\partial\hat{\Gamma}}{\partial\hat{a}}\rangle-2\langle\hat{a}^\dagger\rangle\langle\hat{\Gamma}\rangle\big).
\end{array}
\end{eqnarray}
Applying the coherent state approximation hence yields the
desired classical equations of motion governed by the
complex valued Hamiltonian function $\cH=H-\rmi\Gamma$, where
$H$ and $\Gamma$ are the Weyl symbols of the operators (i.e. the expectation values 
in coherent states):
\begin{equation}
\rmi\dot{z}=\frac{\partial {H}}{\partial z^*}-\rmi\frac{\partial
\Gamma}{\partial z^*}=\frac{\partial \cH}{\partial z^*},\qquad
\rmi\dot{z}^*=-\frac{\partial {H}}{\partial z}-\rmi\frac{\partial
\Gamma}{\partial z}=-\frac{\partial \cH^*}{\partial z}\, . 
\label{cleq-z}
\end{equation}
The evolution equation (\ref{norm}) for the norm is then approximated by
\begin{eqnarray}
\hbar \dot n=-2(\Gamma+\Gamma_0) n,
\label{norm-z}
\end{eqnarray}
with constant $\Gamma_0$ (note that $\Gamma=\Gamma(z,z^*)$ depends implicitly on time). This evolution
equation for the norm in fact goes beyond a purely classical description because
it depends on $\hbar$. Furthermore, as will be become clear from the examples below,
$\Gamma$ tends to zero in the long time limit and the constant 
$\Gamma_0$ accounts for the finite asymptotic decay rate (\ref{Ninf}).

Translated to $q$ and $p$ the equations of motion (\ref{cleq-z}) 
can be expressed as
\begin{equation}
\dot{q}=\frac{\p {H}}{\p p}-\frac{1}{m\omega}\,\frac{\p {\Gamma}}{\p q}\quad {\rm
and}\quad \dot{p}=-\frac{\p {H}}{\p q}-m\omega\frac{\p {\Gamma}}{\p
p}\label{eqn-can-pq-nherm-flat}.
\end{equation}
For the time evolution of a dynamical variable $A(q,p)$ with 
$\p A/\p t=0$ we therefore find
\ba
\dot{A}(q,p)&=&\frac{\p A}{\p q}\dot{q}+\frac{\p A}{\p p}\dot{p}\nn\\
&=& \frac{\p A}{\p q}\frac{\p H}{\p p}-\frac{\p A}{\p p}\frac{\p H}{\p q}-
\frac{1}{m\omega}\,\frac{\p A}{\p q}\frac{\p \Gamma}{\p q}-m\omega\frac{\p A}{\p p}\frac{\p \Gamma}{\p p}\\
&=& \lbrace A,\ H \rbrace_{q,p}-
\left( \frac{1}{m\omega}\,\frac{\p A}{\p q}\frac{\p
\Gamma}{\p q}+m\omega\frac{\p A}{\p p}\frac{\p \Gamma}{\p p}\right).\nn \ea

Equation \rf{eqn-can-pq-nherm-flat} can be represented in a generalised 
canonical structure. To see this we introduce phase space variables 
of equal dimension via the canonical transformation
\begin{equation}
q\to q/\sqrt{m\omega},\qquad p\to\sqrt{m\omega}\,p.
\end{equation}
Then equation \rf{eqn-can-pq-nherm-flat} can be written
in the form
\begin{equation}\label{eqn-grad-flow}
\left(\begin{array}{c}
\dot q\\
\dot p\end{array}
\right)= \symp^{-1}\nabla H-G^{-1}\nabla\Gamma,
\label{genHeq}
\end{equation}
where $\nabla$ is the phase space gradient operator, $\symp$ 
is the symplectic unit matrix and $G$ denotes the phase space 
metric. Since the considered phase space has Euclidean geometry, 
we thus have:
\begin{equation}
\symp=\left(\begin{array}{cc}
0 & -1 \\
1 & 0
\end{array}\right)\quad ,\quad
G=\left(\begin{array}{cc}
1 & 0 \\
0 & 1
\end{array}\right)
\label{eqn-sympl}
\end{equation}
(see appendix \ref{chap-metric} and \cite{09nhbh} for further details). 

The dynamical equation (\ref{eqn-grad-flow}) is a combination of a
canonical symplectic flow generated by the real part $H$ of the
Hamiltonian function and a canonical metric gradient flow generated by the
imaginary part $\Gamma$ of the Hamiltonian function. 
The symplectic part evidently gives rise to
the familiar Hamiltonian dynamics of classical mechanics. The
appearance of the gradient flow may at first seem surprising in the
present context. However, if we recall the fact that the gradient
vector with a negative sign points in the direction of the steepest
descent of the function $\Gamma$ we see that this part of the
dynamics drives the system towards  the minimum of $\Gamma$ and thus
can naturally be associated with a damping 
(see, e.g., \cite{Kauf84,Morr86,Bloc92,Brod96,Holm07}). 
We shall show in section \ref{sec_ang} that 
this structure persists for a spherical phase space arising 
as the classical limit of the quantum angular momentum 
system studied in \cite{08nhbh_s,09nhbh}.

Finally it should be pointed out that the frequency $\omega$ appearing
in the classical equations of motion, i.e. the frequency of the
harmonic oscillator introduced to define the coherent states, is
still a parameter and can be adjusted, for example to improve the
quality of the classical approximation.

In summary we identified a generalised classical canonical structure \rf{genHeq} associated 
with non-Hermitian dynamics for a flat phase space. We derived this structure 
from a coherent state approximation and it should be noted, that it does not 
depend on the specific choice of $\hat H$ and $\hat \Gamma$. In the case where 
both are linear in the generators of the Heisenberg-Weyl algebra, 
$\hat a,\, \hat a^\dagger$ and $\hat a\hat a^\dagger$, and for a coherent initial 
state this approximation becomes exact and the quantum dynamics is 
fully captured by the classical equations of motion. To get further insight into the 
quantum classical correspondence for non-Hermitian systems, we shall 
present some example studies in the following section.

\section{The damped oscillator}
\label{sec-damped-osc} 
In this section we will study the dynamics of a damped oscillator starting
with a purely harmonic case. Here the classical approximation is exact, 
provided that the state is initially coherent. This is a well known 
correspondence identity for Hermitian systems which also extends to the
non-Hermitian case. In the subsequent section we present some results
of a case study of an anharmonic oscillator, where the classical dynamics
is approximate.
\subsection{The damped harmonic oscillator}
Let us first consider the purely harmonic case
\begin{equation}
\label{HO}
\hat H_0=\frac{1}{2m}\hat p^2+\frac{m\omega^2}{2}\hat q^2\, ,
\end{equation}
and assume that the non-Hermitian part of the 
Hamiltonian is 
\begin{equation}
\hat \Gamma =k\hat H_0
\label{gamma}
\end{equation}
with a positive 
real number $k$. Thus, the metric gradient part of the resulting 
dynamics \rf{genHeq} will drive the system towards  the minimum of the 
energy. The classical equations of motion (\ref{eqn-can-pq-nherm-flat}) are
\begin{equation}
\dot{q}=p/m -\gamma q \ , \quad \dot{p}=-m\omega^2q -\gamma p,
\label{classDHO}
\end{equation}
with $\gamma = k\omega$. 
Eliminating the momentum $p$ we obtain
\begin{equation}
\ddot{q}+2\gamma \dot q+(\omega^2+\gamma^2)q=0\,,
\label{classDHOq}
\end{equation}
which is the familiar classical equation of motion for a damped harmonic
oscillator, however, with a frequency $\omega_0=\sqrt{\omega^2+\gamma^2}$,
which implies that the damping is always subcritical. The evolution
equation for the damped harmonic oscillator (\ref{classDHOq}) 
has been presented by Gisin in \cite{Gisi81a}. 

Here the Hamiltonian $\hat {\cal H}$ is linear in the generators
$\hat a$, $\hat a^\dagger$ and $\hat a^\dagger\hat a+1/2$ of the harmonic oscillator 
algebra. Hence an initially coherent state remains coherent
and the classical evolution equation for $q$ and $p$ agree {\it exactly\/} with the quantum
evolution of the expectation values 
$\langle \hat q\rangle$ and $\langle \hat p\rangle$. 

Note, however, that this is no
longer the case if one uses (i) coherent states in the classicalisation
belonging to a harmonic oscillator with frequency $\omega'\ne \omega$; or
(ii) a different damping operator $\hat \Gamma$, for example, 
$\hat \Gamma=k\hat p^2/2m$ (physically that means that the anti-Hermitian part seeks 
to minimise the kinetic energy only). In the first case one obtains the classical
evolution equation
\begin{equation}
\ddot{q}+\gamma \Big( \frac{\omega'}{\omega}+ \frac{\omega'}{\omega}\Big)\dot q+(\omega^2+\gamma^2)q=0\,,
\end{equation}
and in the second case one finds
\begin{equation}
\ddot{q}+2\gamma\dot q+\omega^2q=0\,.
\end{equation}
Both results are now approximations of the true quantum dynamics because the
Hamiltonian is no longer linear in the generators of the algebra. 
The same classical equation of motion also appears in the dynamics 
of a Caldirola-Kanai Hamiltonian which describes the classical dynamics 
of a harmonic oscillator whose mass increases exponentially in time. 
This Hamiltonian can immediately be quantised (see the recent review \cite{Lope08}) 
and yields dynamics different from the one investigated here. 

In what follows, we choose the proper coherent states and $\hat \Gamma$ as
given in (\ref{gamma}). In this case, the Hamiltonian can be simply written
as
\be
\hat \cH=\hbar(\omega-\rmi \gamma)(\hat a^\dagger\hat a+1/2)\,,
\ee
a harmonic oscillator with a complex frequency $\om =\omega-\rmi \gamma$, 
and many results derived for the Hermitian harmonic oscillator are also
valid here. For instance the time evolution of an initially coherent state
$|\alpha_0\rangle$ is given by (see, e.g., \cite{Loui90}) 
\be
|\psi(t)\rangle =\rme^{\rmi \om t/2+D_t\alpha_0\hat a^\dagger}|\alpha_0\rangle
=\rme^{\rmi \om t/2-|\alpha_0|^2(1-\rme^{-2\gamma t})/2}|\alpha_{\,t}\rangle
\label{psih0}
\ee
with
\be
\label{Dt}
D_t=\rme^{-\rmi \om t}-1\,,
\ee
where $\alpha_{\,t}=(1+D_t)\alpha_0=\rme^{-\rmi \om t}\alpha_0$ satisfies 
the (classical!) equation of motion $\rmi \dot \alpha_{\,t}=\om \alpha$.
The time dependence of the norm goes exponentially to zero for long times,
\be
n_t=\langle \psi(t)|\psi(t)\rangle= \rme^{-\gamma t-|\alpha_0|^2(1-\rme^{-2\gamma t})}
\rightarrow \rme^{-\gamma t} \rightarrow 0\,,
\ee
whereas for short times it decays linearly 
according to $n_t\approx 1-\gamma(1+|\alpha_0|^2)t$.
Note that the time derivative of $n_t$ agrees exactly with the classical
equation (\ref{norm-z}) for $\Gamma_0=k\hbar\gamma/2$, which justifies
the \textit{ad hoc} insertion of this correction term.

\subsection{The forced and damped harmonic oscillator}
We now consider one of the most celebrated 
dynamical systems in textbooks, the (classical)
damped harmonic oscillator with  harmonic driving
\be
\ddot q+2\gamma \,\dot q +\omega_0^2\,q=F_0\cos \Omega t\,.
\ee
Here all trajectories approach the limit cycle 
\be
q(t)=Q\cos(\Omega t -\delta),
\label{limit}
\ee
where the amplitude and phase shift are 
\be
Q^2=\frac{F_0^2}{(\omega_0^2-\Omega^2)^2+4\gamma^2\Omega^2}
\,,\qquad \delta=\arctan \frac{2\gamma\Omega}{\omega_0^2-\Omega^2}\,.
\ee
The corresponding quantum system has been studied only rarely (see, e.g.,
\cite{Gisi83a,92dqs}). Here we consider the Hamiltonian
\be
\cH=\hbar \om (\hat a^\dagger\hat a+1/2)+\hbar f_t\hat a + \hbar f^*_t\hat a^\dagger\,,
\ee
again with a complex frequency $\om$ and a time dependent 
driving $f_t$. The closed form solution known
for real frequency (see, e.g., \cite[§ 3.11]{Loui90}) 
can be extended to the non-Hermitian case
with complex frequency $\om $. For convenience we provide the solution here
for an initially coherent state $|\alpha_0\rangle$:
\begin{eqnarray}
\fl |\psi(t)\rangle =\rme^{-\rmi \om t/2+A_t+B_t\alpha_0+(C_t+D_t\alpha_0)\hat a^\dagger}|\alpha_0\rangle
=\rme^{-\rmi \om t/2+A_t+B_t\alpha_0-|\alpha_0|^2/2+|\alpha_{\,t}|^2/2}
|\alpha_{\,t}\rangle,
\label{psihf}
\end{eqnarray}
with $\alpha_{\,t}=C_t+(1+D_t)\alpha_0$. The parameters satisfy the
differential equations
\be
\fl \quad
\rmi \dot D=\om (D+1)\,,\quad 
\rmi \dot B=f_t(D+1)\,,\quad 
\rmi \dot C=\om C +f^*_t\,,\quad 
\rmi \dot A=f_tC
\ee
with initial conditions $A_0=B_0=C_0=D_0=0$. The solutions are
\begin{eqnarray}
\fl \quad B_t=-\rmi \int_0^t\!\!\rmd t'\rme^{-\rmi \om t'}f_{t'}\ ,\quad 
C_t=-\rmi \int_0^t\!\!\rmd t'\rme^{\rmi \om (t'-t)}f^*_{t'}\ ,\quad  
A_t=-\rmi \int_0^t\!\!\rmd t'f_{t'}C_{t'}
\end{eqnarray}
and $D_t$ is again given by (\ref{Dt}).
Note that $\alpha_{\,t}$, parametrising the coherent state, is a solution of
the differential equations
\be
\rmi \dot \alpha=\om \alpha +f_t^*\,,\quad \rmi \dot \alpha^*=-\om^* \alpha^* -f_t,
\ee
which are indeed the classical equation of motion
\be
\rmi \dot \alpha=\frac{\partial \cH}{\partial \alpha^*}\,,\quad 
\rmi \dot \alpha^*=-\frac{\partial \cH^*}{\partial \alpha}
\ee
for the Hamiltonian $\cH = \om \alpha^*\alpha +f_t\alpha +f_t^*\alpha^*$, where
$\alpha$ and $\alpha^*$ act as canonical coordinate and momentum, respectively.
For the special case of harmonic driving, $f_t=f_0\cos \Omega t$, one obtains
\begin{eqnarray}
\fl \quad B_t&=& \frac{f_0}{\om^2-\Omega^2}\Big\{(\om+\Omega)\rme^{-\rmi (\om -\Omega)t}
+(\om-\Omega)\rme^{-\rmi (\om +\Omega)t}-2\om\Big\},\\
\fl \quad C_t&=&  -\frac{f_0}{2(\om^2-\Omega^2)}\Big\{(\om-\Omega)\rme^{\rmi \Omega t}
+(\om+\Omega)\rme^{-\rmi \Omega t }-2\om\rme^{-\rmi \om t}\Big\},\\
\fl \quad A_t&=&  \frac{f_0^2}{4(\om^2-\Omega^2)}\Big\{2\rmi\om t +\frac{1}{2\Omega}
\big((\om -\Omega)\rme^{2\rmi \Omega t} - (\om +\Omega)\rme^{-2\rmi \Omega t} +2\Omega\big)\nn\\
\fl && \qquad \qquad +\frac{2\om}{\om^2-\Omega^2}\big((\om +\Omega)\rme^{-\rmi (\om- \Omega) t} 
+ (\om -\Omega)\rme^{-\rmi(\om+ \Omega) t} -2\om\big)\Big\},
\end{eqnarray}
which approach, in the long time limit, to the simpler expressions $D_t  \rightarrow\ -1$, $B_t  \rightarrow\ 0$,
$A_t\rightarrow A^\infty_t+A^{(0)}_t$, with 
\begin{eqnarray}
 && A^\infty_t=\frac{\rmi f_0^2}{2(\om^2-\Omega^2)}\om t,\\
&& A^{(0)}_t= \frac{f_0^2}{4(\om^2-\Omega^2)}\Big\{\frac{1}{2\Omega}
\big((\om -\Omega)\rme^{2\rmi \Omega t} - (\om +\Omega)\rme^{-2\rmi \Omega t} 
\big)\Big\}\, ,
\end{eqnarray}
and
\begin{eqnarray}
\fl &&C_t \rightarrow -\frac{f_0}{2(\om^2-\Omega^2)}\Big\{(\om-\Omega)\rme^{\rmi \Omega t}
+(\om+\Omega)\rme^{-\rmi \Omega t }\Big\}=\alpha_{\,t}^{(0)}
\end{eqnarray}
with $\alpha_{\,t} \rightarrow \alpha^{(0)}_{\,t}$, the limit cycle (\ref{limit}).
We note that $\alpha^{(0)}_{\,t}$ and $A^{(0)}_t$ are $T$-periodic  and 
the limiting state for $t\rightarrow \infty$ can be written in the form
\be
\fl \ |\psi_0(t)\rangle = \rme^{-\rmi \epsilon_0 t/\hbar}|\phi_0(t)\rangle 
\quad \textrm{with}\quad 
|\phi_0(t)\rangle = c\rme^{-\rmi A^{(0)}_t+|\alpha_{\,t}|^2/2} |\alpha^{(0)}_{\,t}\rangle
=|\phi_0(t+T)\rangle\,,
\label{psi0}
\ee
where $c$ is a constant, and 
\be
\epsilon_0=\frac{\hbar \om}{2}\Big(1-\frac{f_0^2}{\om^2-\Omega^2}\Big)\,.
\label{epsilon0}
\ee

\begin{figure}[h]
\centering
\includegraphics[height=5cm,  angle=0]{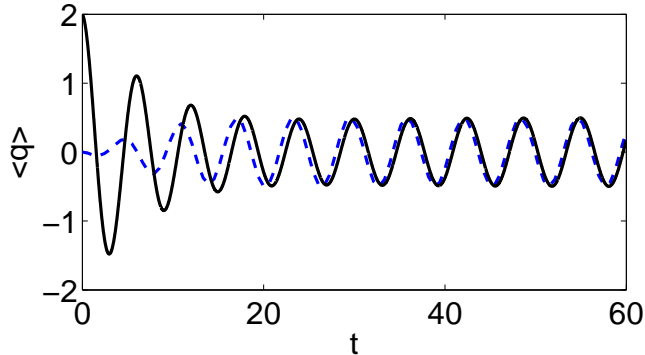}
\caption{\label{fig-cat} Time evolution
of the mean position $\langle \hat q\rangle$
for a damped quantum harmonic oscillator ($\omega=m=\hbar=1$, $\gamma=0.1$)
which is harmonically driven ($f_0=0.1$, $\Omega=1$). The solid black 
curve shows the evolution of a coherent state initially located
at $(q_0,p_0)=(2,0)$. This curve agrees with the classical motion and converges
to the limit cycle. The dashed blue curve is obtained for an initial superposition
of two coherent states located at $(2,0)$ and $(-2,0)$.}
\end{figure}

The quantum dynamics can be computed numerically, e.g., in the
discrete operator representation \cite{02computing}. 
As an example, figure \ref{fig-cat} shows the evolution of a coherent 
state initially located at $(q_0,p_0)=(2,0)$.
The parameters are chosen as $\omega=1$, $m=1$
(these are kept fixed in all calculations) and $\gamma=0.1$, $\hbar=1$. 
For the driving force we used $\Omega=1$ and $f_0=0.1$.
The time dependence of the expectation value $\langle \hat q\rangle$ 
shown as a solid black curve agrees, of course, with the analytic formula 
given above and also with the classical oscillation $q(t)$. 
Asymptotically, the motion approaches the limit cycle (\ref{limit}).

If the initial state is chosen differently from a coherent one, 
the agreement between classical and quantum
evolution breaks down, as in the Hermitian case. As an example, the dashed blue curve
in the figure shows the results for an initial superposition of two 
coherent states located at $(q_0,p_0)=(2,0)$  and  $(-2,0)$ (a so-called cat state).
Here $\langle \hat q\rangle$, which is initiallly equal to zero, cannot
be described by a single classical trajectory but shows an interference structure
before it converges to the asymptotic limit state. This behaviour is characteristically different 
from the Hermitian case, where no limit cycle exists and the quantum 
evolution never approaches the classical dynamics.  

Let us finally remark that in the quantum case most states approach asymptotically 
to a unique state, a quantum limit cycle. This is evident, if one formulates
the dynamics in terms of non-Hermitian Floquet states. Here the time evolution 
can be expressed in the form of a linear combination 
as in (\ref{psit-n}), where the eigenstates 
are replaced by the Floquet states 
\be
|\psi_n(t)\rangle = \rme^{-\rmi \epsilon_nt/\hbar}|\varphi_n(t)\rangle
\quad \textrm{with}\quad |\varphi_n(t+T)\rangle =|\varphi_n(t)\rangle
\ee
and the eigenvalues by the (complex) quasienergies $\epsilon_n$. 
Then each state approaches asymptotically the most stable Floquet state, which can
be identified as the quantum limit cycle. In the present case of a forced harmonic
oscillator the quasienergies are known analytically (see, e.g., \cite{ditt92,02computing}) 
and the solution can be directly extended to complex frequencies:
\be
\epsilon_n=\hbar \om \,\Big(n+\frac{1}{2}-\frac{f_0^2}{2(\om^2-\Omega^2)}\Big)\ ,
\quad n=0,\,1,\,\ldots
\ee
which agrees for $n=0$ with the result already given in (\ref{epsilon0}). 
The most stable state $n=0$ is, of course, only reached if the 
projection of the initial state onto this state is different from zero. 
Otherwise, the state approaches the most stable initially populated 
quasienergy state. In fact, all quasienergy states appear as quantum limit cycles,
as already remarked by Gisin \cite{Gisi83a}.

\begin{figure}[h]
\centering
\includegraphics[height=7cm,  angle=0]{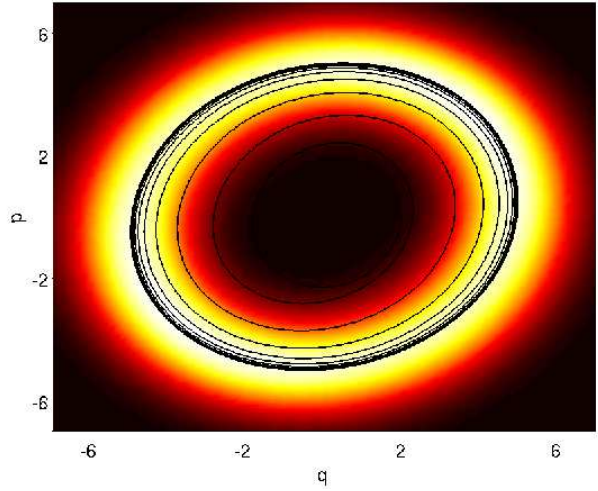}
\caption{\label{fig-husimi} Husimi distribution of the quantum 
limit cycle for a damped quantum harmonic oscillator 
($\omega=m=\hbar=1$, $\gamma=0.1$)
which is harmonically driven ($f_0=1$, $\Omega=1$), averaged 
over one driving period. The solid black line shows a classical trajectory 
started at $(q_0,p_0)=(2,0)$ for comparison. 
This curve approaches the classical limit cycle that coincides with the 
ridge of the averaged Husimi distribution of the quantum limit cycle.}
\end{figure}
As an illustration figure \ref{fig-husimi} shows an example of the quantum limit cycle in 
phase space. Shown in false colors is the Husimi distribution, i.e.~the projection on 
coherent states
\be
P(p,q;t)=|\langle \alpha|\psi(t)\rangle|^2\  ,\quad \alpha=(m\omega q+\rmi 
p)/\sqrt{2m\hbar\omega}
\ee
which is averaged over one period of the driving in the long time limit. Also shown is a classical 
trajectory started in at $p_0 = 0$, $q_0 = 2$ which 
converges to the classical limit cycle that concides with the ridge of the
quantum distribution.

\begin{figure}[b]
\centering
\includegraphics[height=5cm,  angle=0]{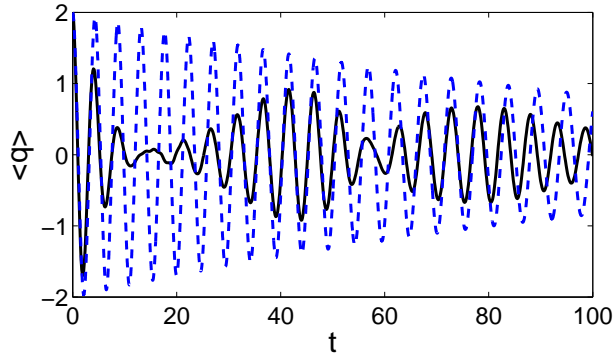}
\caption{\label{fig-revival} Comparison of the quantum and classical
damped anharmonic oscillator with $\beta=0.4$ and weak damping, $\gamma=0.01$.  
The quantum expectation value $\langle q\rangle$ (solid black curve) is shown as 
a function of time in comparison with the
classical evolution $q(t)$ (dashed blue curve).}
\end{figure}
\subsection{A damped anharmonic oscillator}
As an example of an anharmonic oscillator we consider the Hamiltonian
\begin{equation}
\label{AHO}
\hat H=\frac{1}{2m}\hat p^2+\frac{m\omega^2}{2}\hat q^2+\frac{\beta}{4}\hat q^4\,,
\end{equation}
where the non-Hermitian part $\hat \Gamma$ of the Hamiltonian is chosen as
\begin{equation}
\hat \Gamma=\frac{\gamma}{\omega}\left(\frac{1}{2m}\hat p^2+\frac{m\omega^2}{2}\hat q^2\right), 
\end{equation} 
as for the harmonic oscillator. Then
the classical equations of motion (\ref{eqn-can-pq-nherm-flat}) are
\begin{equation}
\dot{q}=p/m -\gamma q \ , \quad \dot{p}=-m\omega^2q-\beta q^3 -\gamma p
\label{classDAHO}.
\end{equation}
In the Hermitian case $\gamma=0$ one finds the well-known deviations 
between the quantum and classical evolution showing breakdown and revival
phenomena. As illustrated in figure \ref{fig-revival}, these phenomena survive
in the case of a weak damping,
$\gamma=0.01$, and moderate anharmonicity $\beta=0.4$. The oscillation frequency
and the overall decay is reproduced in the classical approximation. 

\begin{figure}[t]
\centering
\includegraphics[height=3cm,  angle=0]{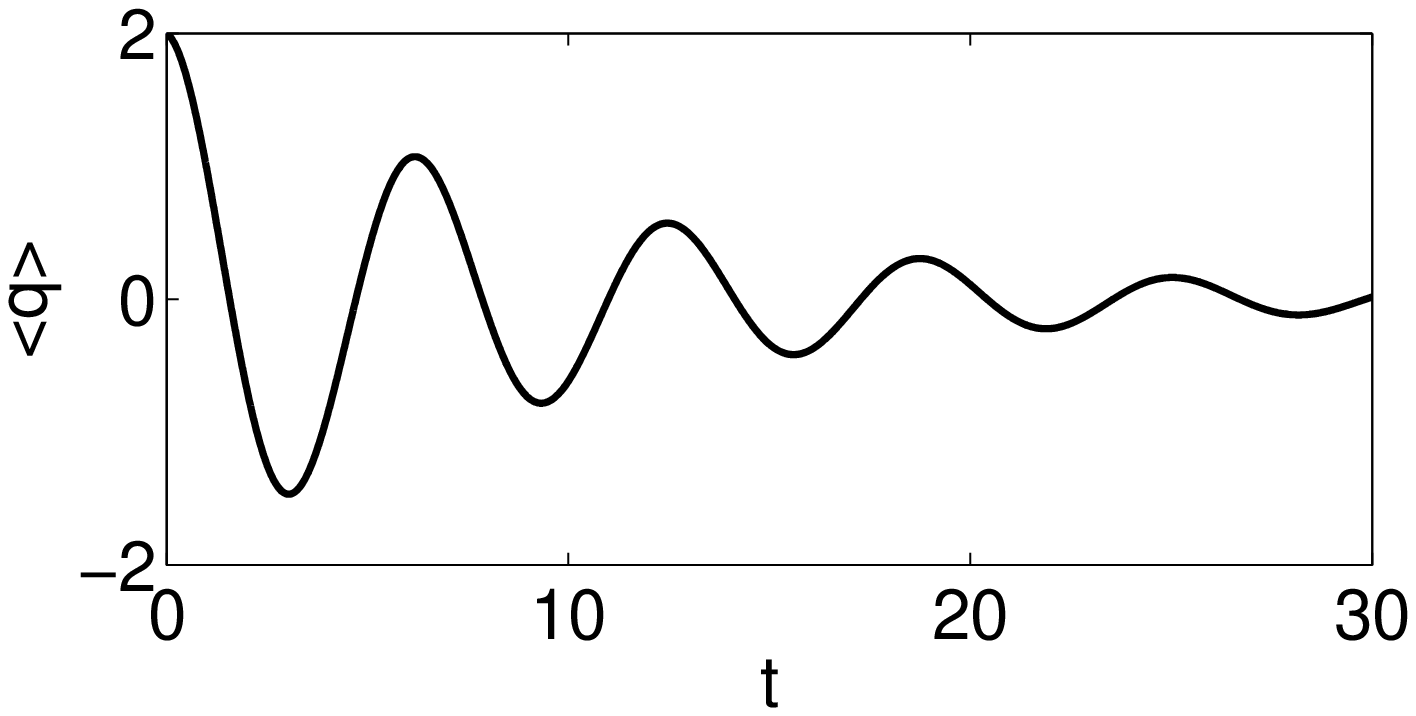}
\includegraphics[height=3cm,  angle=0]{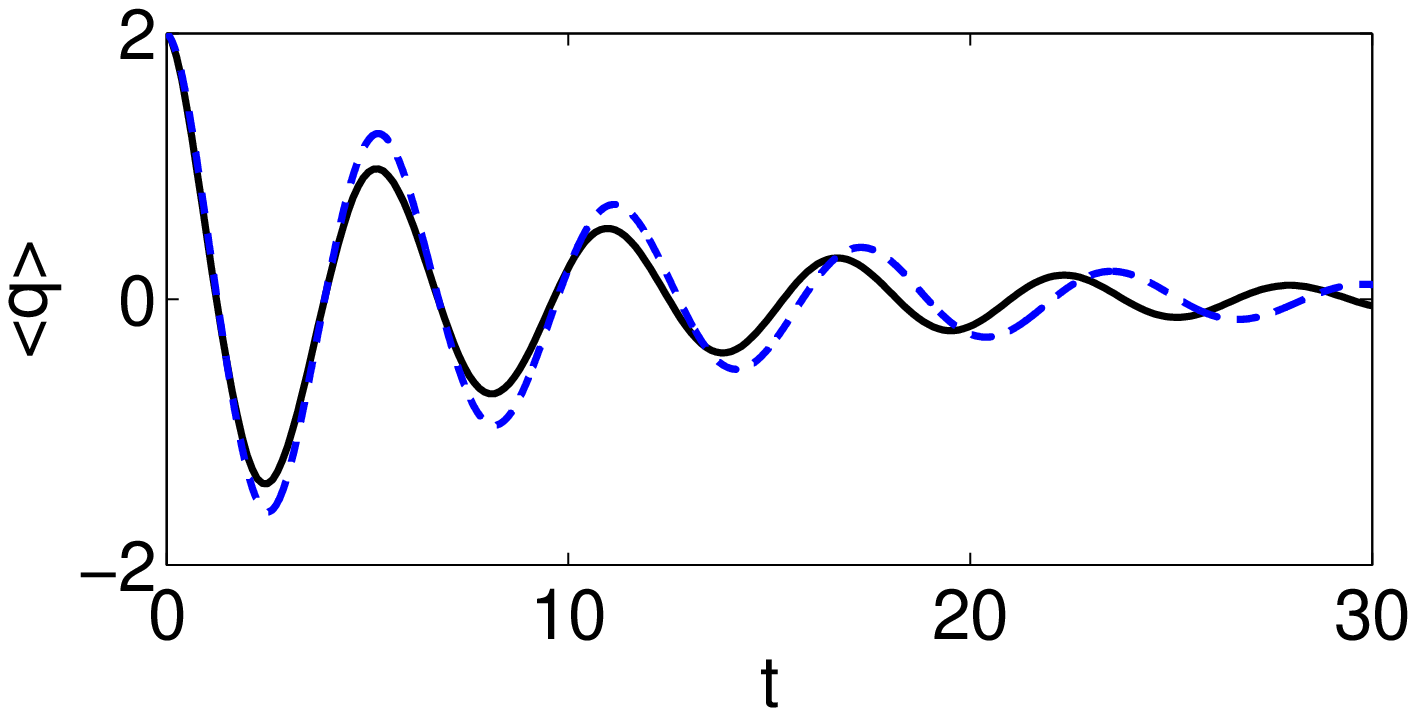}
\includegraphics[height=3cm,  angle=0]{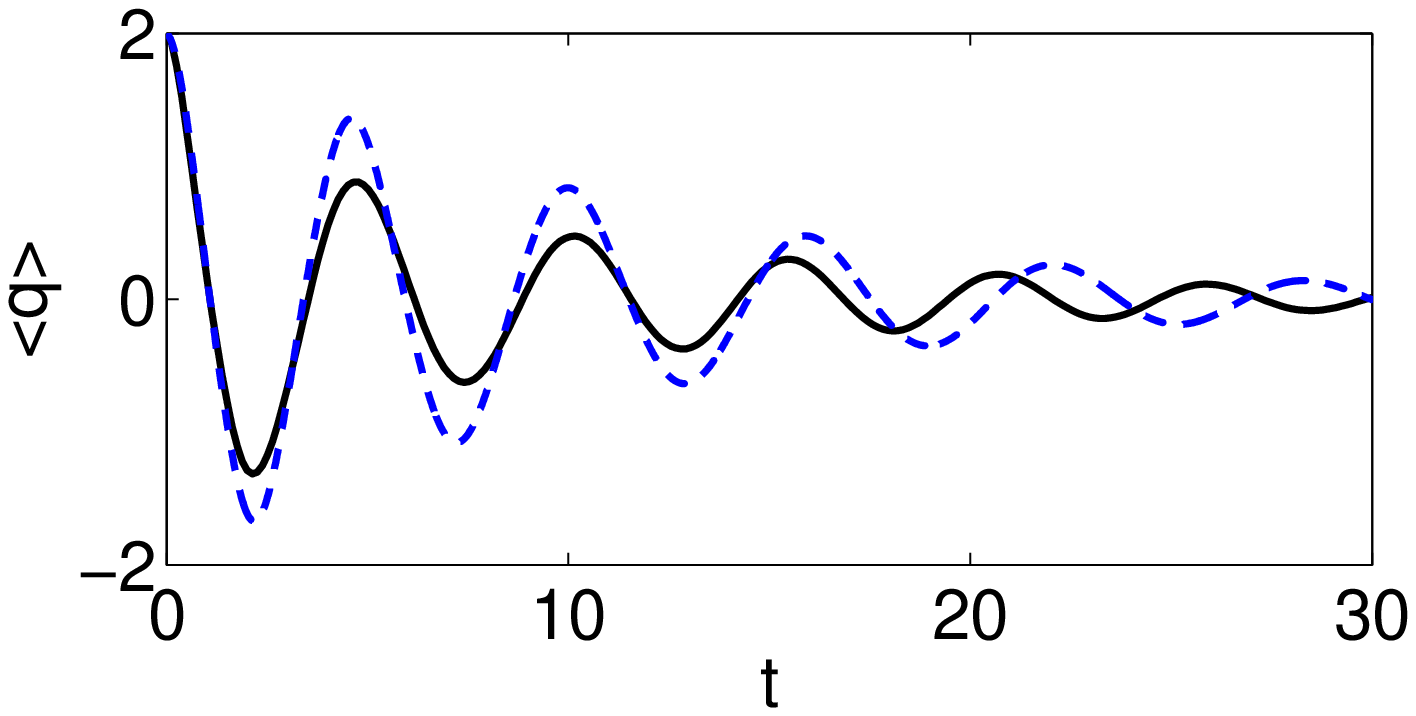}
\includegraphics[height=3cm,  angle=0]{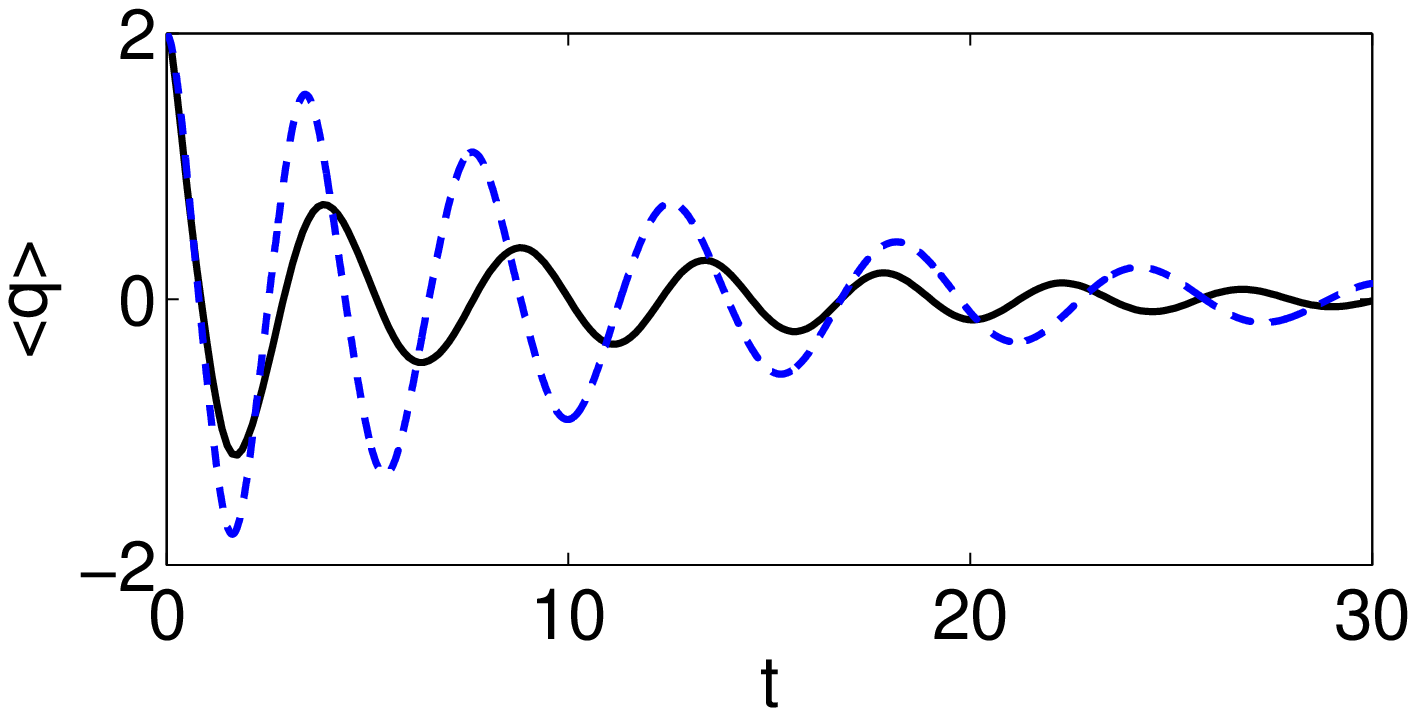}
\caption{\label{fig-anh_beta} Damped harmonic and
anharmonic oscillator: 
Expectation value $\langle q\rangle$ as 
a function of time for $\gamma=0.1$ and $\beta=0$ (upper left)
$\beta=0.2$ (upper right), $\beta=0.4$ (lower left) and
$\beta=1.0$ (lower right). The quantum expectation values 
$\langle q\rangle(t)$ (black solid curves) for $\hbar=1$ are compared with the
classical evolution $q(t)$ (blue dashed curves).}
\end{figure}

Numerical results for the damped case with stronger damping are shown in figure \ref{fig-anh_beta},
where we have used the same method and parameters as described above.
The damped anharmonic evolution shows differences between the
classical and quantum evolution and, as expected, the deviations
increase with anharmonicity $\beta$. This is due to the increasing
deviation of the quantum state from a coherent one during the
time evolution. For short times, when the state is still almost coherent,
we find a very good agreement between classical and quantum evolution.
In further numerical studies we observed that the qualitative behaviour 
is not changed, if we also chose the anti-Hermitian part $\hat \Gamma$ 
anharmonic. A detailed investigation of the behaviour of the 
quantum-classical correspondence in dependence on the special choice of 
Hermitian and anti-Hermitian part as well as their relation 
is a promising starting point for future studies.

\section{An angular momentum system}
\label{sec_ang}
Recently a generalised classical approximation in the spirit 
introduced in the preceding sections has been performed for a non-Hermitian angular momentum 
system  in the context of many-particle mean-field correspondence 
for ultracold atoms in an open double well trap \cite{08nhbh_s,09nhbh}. 
The Hamiltonian investigated there is of the form 
\begin{eqnarray}
\hat{\cal H}=2 \epsilon\hat L_z+2\J \hat L_x+2\U \hat
L^2_z-2\rmi \gamma (\hat L_z+L) \,, \label{BH-hamiltonian-L}
\end{eqnarray}
where the $\hat L_j$ are angular momentum operators and $L$ denotes the 
angular momentum quantum number. 
By Schwinger's harmonic oscillator representation of angular momentum, this 
can be interpreted as a second quantised many-particle model and the 
classical approximation arises as a mean-field approximation. Here we briefly review 
the generalised coherent state approximation for this angular momentum system and 
show that the resulting mean-field dynamics can be described by the generalised 
canonical structure \rf{genHeq}, however, on a spherical phase space. 

To perform the classical approximation we have to replace expectation values 
in the quantum Heisenberg equations of motion with their values in a coherent state. 
For angular momentum operators the classical approximation will 
be performed using $SU(2)$ coherent states. The classical limit is 
then realised for large quantum numbers, that is, $L\to \infty$, where we set  
$\hbar=1$ for convenience. 

The dynamics of the angular momentum expectation values follows from 
the generalised Heisenberg equation of motion \rf{GHE}:
\begin{eqnarray}\label{KomZerfall1}
{\textstyle \frac{{\rmd}}{\rmd\, t}} \langle \hat L_x \rangle &=&
- 2 \epsilon \langle \hat L_y \rangle - 2\U  \langle [ \hat L_y, \hat L_z]_{\scriptscriptstyle +} \rangle
- 4\gamma  \, \Delta^2_{\hat L_x\hat L_z}   \nonumber\\
{\textstyle \frac{\rmd}{\rmd\, t}} \langle \hat L_y \rangle &=&
 2\epsilon \langle \hat L_x \rangle + 2\U  \langle [ \hat L_x, \hat L_z]_{\scriptscriptstyle +} \rangle
-2\J  \langle \hat L_z \rangle - 4\gamma  \, \Delta^2_{\hat L_y\hat L_z} \\
{\textstyle \frac{\rmd}{\rmd\, t}} \langle \hat L_z \rangle &=&
 2 \J  \langle \hat L_y \rangle - 4 \gamma  \,\Delta^2_{\hat L_z\hat L_z} ,
 \nn
\end{eqnarray}
and the norm of wave function decays according to
\begin{equation}
\frac{\rmd}{\rmd\, t} \langle \Psi| \Psi \, \rangle = -4\gamma
\,\big\lbrace \langle \hat L_z \rangle+ L
\big\rbrace \langle \Psi | \Psi \, \rangle.
\end{equation}
The generalised $SU(2)$ coherent
states, often also denoted as atomic coherent states, 
can be constructed by an arbitrary $SU(2)$ rotation of the
extremal angular momentum state $|L\rangle$ (spin up):
\begin{equation}
\label{eqn-SU2-cs}
|\theta,\phi\rangle =\hat R(\theta,\phi)|L\rangle= {\rm e}^{{\rm i}\theta(\hat L_x \sin{\phi} - \hat L_y\cos{\phi})}|L\rangle.
\end{equation}
The coherent state approximation---that is, the assumption that an initially coherent 
state stays coherent throughout the time evolution---is exact if the Hamiltonian is 
linear in the generators of the $SU(2)$ algebra, i.e. for $c=0$ in our case. 
This can be seen by an explicit calculation of the action of 
the time evolution operator
\begin{equation}
U(t)=\exp{(-\rmi \hat H t/\hbar)}
\end{equation}
on an initially coherent state \rf{eqn-SU2-cs}. 
The classical equations of motion are obtained from the quantum dynamics of the 
relevant expectation values by replacing all expectation values with their 
values in coherent states and identify these as the classical quantities. 
The $SU(2)$ expectation values of the $\hat L_j$, $j=x, y, z$ appear as
the components of the classical Bloch vector:
\begin{equation}
\s_j=\langle \hat L_j \rangle/2L.
\end{equation}
The expectation values of the anti-commutators appearing 
in \rf{KomZerfall1} in $SU(2)$ states can be shown \cite{08nhbh_s,09nhbh} to factorise as
\begin{eqnarray} \label{ErwAntiDrehexakt}
\langle [\hat L_i, \hat L_j]_{\scriptscriptstyle +} \rangle = 2
(1-\frac{1}{2L} ) \langle \hat L_i \rangle \langle \hat L_j
\rangle + \delta_{ij} L\, 
\end{eqnarray}
Inserting these expressions into (\ref{KomZerfall1}) and
taking the classical limit $L\to\infty$ (where we have to keep $2L\U =\g $ 
fixed) we obtain the desired non-Hermitian 
classical evolution equations:
\begin{eqnarray}\label{eqn-bloch-nherm-nlin}
\begin{array}{rrrrll}
\dot{\s_x}=&-2 \epsilon \s_y&- 4 \g \s_z \s_y& &+ 4 \gamma \,\s_z
\s_x, \\ \dot{\s_y}=&+2 \epsilon \s_x&+4 \g \s_z \s_x &-2 \J  \s_z
&+ 4 \gamma \,\s_z \s_y,\\ \dot{\s_z}=& & &+2 \J  \s_y&-\gamma\,
(1-4\s_z^2)\,.
\end{array}
\end{eqnarray}
In the case $g=0$ in which the assumption that the many-particle
state stays coherent in time is exactly fulfilled, these equations
exactly coincide with the quantum dynamics for arbitrary $L$ for 
an initially coherent state. 
The nonlinear non-Hermitian Bloch equations
are real valued, and conserve $\s^2=\s^2_x+\s^2_y+\s^2_z=1/4$, i.e.~the dynamics is
regular and confined to the Bloch sphere. The decay of the 
total probability can be approximated by 
\begin{equation}
\dot n = -4\gamma L(2s_z+1)n
\end{equation}

The classical dynamics is analysed in detail in \cite{08nhbh_s,09nhbh}. Here 
we show that it can be described by the generalised canonical structure \rf{genHeq}. 
For this purpose we introduce the canonical coordinates 
$p$ and $q$ on the sphere related to the Bloch coordinates via
\begin{eqnarray}
\s_x&=&\half\sqrt{1-p^2}\cos(2q)\nn\\
\s_y&=&\half\sqrt{1-p^2}\sin(2q)\\
\s_z&=&\half p.\nn
\end{eqnarray}
The equations of motion then read:
\begin{eqnarray}
\dot{q}&=&\epsilon+gp-v\frac{p}{\sqrt{1-p^2}}\cos(2q)\\
\dot{p}&=&-2\gamma(1-p^2)+2v\sqrt{1-p^2}\sin(2q).
\end{eqnarray}
This can be expressed as generalised Hamiltonian equations of motion 
of the form (\ref{genHeq}) where $\symp$
is the usual symplectic matrix (\ref{eqn-sympl}) and
\begin{equation}\label{eqn-metric-Bloch-pq}
G=\left(\begin{array}{cc}
2(1-p^2) & 0\\
0 & \frac{1}{2(1-p^2)}
      \end{array}\right)
\end{equation} 
is the corresponding K\"ahler metric on the Bloch sphere 
(see appendix \ref{chap-metric} for some details).
The Hamiltonian function $\cH=H-\rmi\Gamma$ is given by the
$SU(2)$ expectation value of the quantum Hamiltonian
\rf{BH-hamiltonian-L}:
\begin{eqnarray}\label{eqn-nherm-nlin-ham-fct}
H=\epsilon p+v\sqrt{1-p^2}\cos(2q)+\frac{g}{2}p^2\quad{\rm and}\quad
\Gamma=\gamma p.
\label{classical-E-pq}
\end{eqnarray}
\begin{figure}[t]
\begin{center}
\includegraphics[width=8cm]{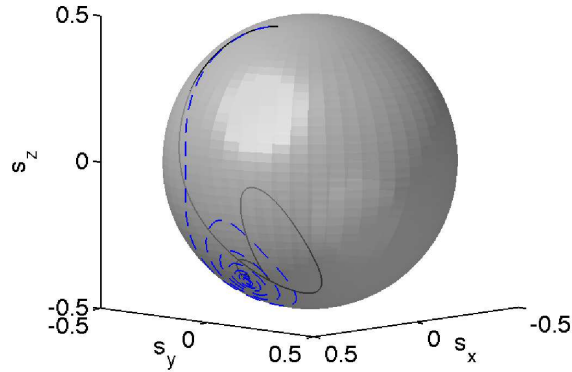}
\caption{\label{fig-bloch} Quantum (black solid line) and classical (blue dashed line) 
angular momentum evolution on the Bloch sphere. Plotted are the expectation values of the 
quantum angular momentum components in comparison with the classical Bloch vector for 
an initially coherent state centered at the classical initial condition close to the 
north pole of the Bloch sphere up to a time $t_end=25$ for $\gamma=0.1$ and $g=1.5$ and 
a total angular momentum of $L=40$.}
\end{center}
\end{figure}

In figure \ref{fig-bloch} we show an example of the quantum dynamics 
in comparison with the classical evolution on the Bloch sphere for 
a strong nonlinearity $g=1.5$ and a small non-Hermiticity $\gamma=0.1$. 
Here the classical dynamics has several fixed points, 
one of them being a sink of the dynamics close to the south pole of the sphere and 
another being a source close to the north pole. The initial condition in the present 
example is close to the source. For short times the behaviour of the quantum and 
the classical dynamics is similar. For longer times, however, while 
the classical dynamics moves towards the sink staying confined 
on the surface of the Bloch sphere, the quantum 
dynamics can tunnel through the sphere. Finally they both approach the 
sink of the dynamics. Note that the quantum classical correspondence 
for this system can be quite intricate. Depending on the initial conditions 
and the total angular momentum the quantum dynamics can differ considerably 
from the underlying classical behaviour due to interference and tunneling effects.  
This is analysed in more detail in \cite{09nhbh}.

\section{Conclusions and Outlook}
In conclusion we derived a generalised canonical structure 
\rf{genHeq} incorporating a 
metric gradient flow that can describe both conservative and dissipative 
motion as a classical limit of non-Hermitian 
quantum dynamics. The classical limit was defined as a coherent state approximation 
and it was shown that the canonical structure arises for Euclidean phase space 
geometry associated with Glauber coherent states, irrespectively of the special 
choice of the Hamiltonian. Furthermore,  we presented 
an example of an angular momentum system where the classical limit 
is given by the limit of large angular momentum. The appropriate coherent 
states in this context are the $SU(2)$ or atomic coherent states which 
lead to a spherical classical phase space. We showed that the resulting classical 
equations of motion can also be described using the generalised canonical structure 
derived for the Euclidean case. This strongly suggests that the proposed structure 
holds for more general systems with different metric structures on the classical phase space.

The quantum classical correspondence arising from these 
non-Hermitian systems is in many cases equivalent 
to the well-known behaviour observed in Hermitian systems. 
However, new effects due to the gradient part of the 
motion are to be expected, the investigation of which is a 
promising task for future studies. Interesting questions here regard 
the timescales of the quantum classical correspondence (that is 
the non-Hermitian equivalents of the 
Ehrenfest and Heisenberg times for Hermitian systems) and their 
dependence on the Hermitian and anti-Hermitian part individually as well 
as the relation of Hermitian and anti-Hermitian part. It is also 
interesting how the generated classical flow depends on this relation 
between the Hermitian and anti-Hermitian part of the Hamiltonian and 
in particular, which simplifications are to be expected if they are proportional. 

Furthermore, the discovery of a general structure in the classical limit of 
non-Hermitian quantum theories paves the way for 
the development of new semiclassical techniques  such 
as quantisation conditions and trace formulas for these systems. 

\section*{Acknowledgments} 
We would like to thank Dorje Brody for inspiring discussions 
and helpful comments. Support from the Deutsche Forschungsgemeinschaft
via the Graduiertenkolleg  ``Nichtlineare Optik und Ultrakurzzeitphysik'' 
is gratefully acknowledged.

\appendix
\section{The Metric, the symplectic structure, and a generalised canonical evolution on phase spaces} 
\label{chap-metric}
In this appendix we provide the minimal basic facts concerning 
the notion of a K\"ahler structure on manifolds that is 
relevant for the generalised canonical structure of 
the non-Hermitian dynamics. The interested
reader is referred to \cite{Arno78,Brod01} for further detail.

In Riemannian geometry a manifold is equipped with a measure of
distance determined by the metric tensor $g$ in the following way:
The infinitesimal distance $\rmd s$ between a point
$(x^1,\cdots,x^n)$ and $(x^1+\rmd x^1,\cdots,x^n+\rmd x^n)$ on an
$n$-dimensional manifold $X_n$ is given by the expression
\begin{equation}
\rmd s^2 = g_{ab}\, \rmd x^{a}\rmd x^{b}.
\end{equation}
Here we make use of the Einstein sum convention. The metric tensor thus
consists of $n(n+1)/2$ quantities $g_{ab}$ that
define the notion of distance on $X_n$.

Clearly distance is invariant under local coordinate
transformation of the form $(x^1,\cdots,x^n)\to (u^1,\cdots,u^n)$.
In other words, if we write $g^{(u)}_{ab}$ for the metric tensor
in the transformed $(u^1,\cdots,u^n)$ coordinates we have
\begin{eqnarray}
\rmd s^2 = g^{(u)}_{ab}\,\rmd u^{a}\rmd u^{b} =  g_{ab}\, \rmd
x^{a}\rmd x^{b},
\end{eqnarray}
from which it follows that the metric transforms according to
\begin{equation}
\label{eq:metric_transform}
g^{(u)}_{ab}=\frac{\partial x^{c}}{\partial u^{a}}\frac{\partial
x^{d}}{\partial u^{b}}\,g_{cd}.
\end{equation}

If the manifold in question is a classical phase space then it
comes equipped with an additional symplectic structure $\omega$. A
space accommodating both a metric $g$ and a symplectic structure
$\omega$ is said to possess a K\"ahler structure if these two
quantities satisfy the compatibility condition \cite{Brod01}, 
which in matrix form is written
\begin{equation}
\omega^{-1}=\big(g^{-1}\omega g^{-1}\big)^{\rm T}, \label{eq-comp}
\end{equation}
where the superscript ${\rm T}$ denotes matrix transpose. It is
evident from (\ref{eq-comp}) that the compatibility condition is
invariant under the simultaneous scale transformation $g\to\kappa
g$ and $\omega\to\kappa\omega$ for some $\kappa\neq0$. Hence, the 
K\"ahler metric and symplectic structure are defined only up to a scale factor. On
the other hand, if either $\omega$ or $g$ is given, then this
automatically fixes the relevant scale factor according to the
compatibility condition.

In classical mechanics one usually associates the symplectic
structure on the phase space with the canonical equations of
motion by demanding them to read
\begin{equation}
\left(\begin{array}{c}
\dot q\\
\dot p\end{array} \right)=
\left(\begin{array}{c}
+\p H/\p p\\
-\p H/\p q\end{array} \right)
= \symp^{-1}\vec{\nabla}H,
\end{equation}
and hence fixing the symplectic structure $\symp$ as
\begin{equation}
\label{eqn-symp-can}
\symp=\left(\begin{array}{cc}
0 & -1 \\
1 & 0
\end{array}\right).
\end{equation}
Thus, if the phase space is a Riemannian manifold and therefore
equipped with a metric $g$ the corresponding K\"ahler metric $G$
is connected to $g$ up to a factor that is determined by the
choice (\ref{eqn-symp-can}) of the symplectic structure. It is
straightforward to show that for a Euclidean space this coincides
with the usual choice of a Euclidean metric $G={\mathds 1}$.

While in the conventional context of Hamiltonian dynamics the
metric is of no consequence, in the present paper we have a 
complex Hamiltonian function $\cH=H-\rmi \Gamma$, and 
the metric structure appears as a natural
choice to extend the canonical equations of motion to the
generalised form
\begin{equation}
\left(\begin{array}{c}
\dot q\\
\dot p\end{array} \right)= \symp^{-1}\vec{\nabla}H-
G^{-1}\vec{\nabla}\Gamma,
\end{equation}
where $\symp$ is given by the usual expression \rf{eq:symp1} for
Hamiltonian dynamics and $G$ is the associated K\"ahler metric.

Let us now proceed to calculate the K\"ahler metric that is
compatible with the choice \rf{eqn-symp-can} for the symplectic
structure, in the case of the Bloch sphere. 
The Riemannian metric $g$ on a two-sphere $S^2
\subset {\mathds R}^3$ of radius $R$, embedded in 
the three dimensional Euclidean phase space via
\begin{eqnarray}
x&=&R\sin\theta\cos\phi\nn\\
y&=&R\sin\theta\sin\phi\\
z&=&R\cos\theta,\nn
\end{eqnarray}
 can be found from the transformation rule \rf{eq:metric_transform} (which extends to the case in
which we have a parametric subspace).
Starting from the metric in ${\mathds R}^3$ given by the identity matrix 
we deduce at once the well-known expression of the metric on the sphere:
\begin{equation}
g^{(\theta,\phi)}=\left(\begin{array}{cc}
R^2 & 0\\
0 & R^2\sin^2\theta
\end{array}\right).
\end{equation}
Because the sphere is the Bloch sphere the
radius is given by $R=\frac{1}{2}$. If we employ the usual
coordinate system $p$ and $q$ according to
\begin{eqnarray}
p=\cos\theta \quad {\rm and} \quad q = \phi/2,
\end{eqnarray}
that is,
\begin{eqnarray}
\theta=\arccos{p} \quad {\rm and} \quad \phi=2q,
\end{eqnarray}
then we find that the metric of our Bloch sphere is given by
\begin{equation}
g^{(p,q)}=\frac{1}{4} \left(\begin{array}{cc}
\frac{1}{1-p^2} & 0\\
0 & 4(1-p^2)
\end{array}\right).
\end{equation}
From this metric we can calculate the symplectic structure
$\omega$ on the phase space $(p,q)$ making use of the K\"ahler
compatibility condition and find:
\begin{equation}
\omega=\left(\begin{array}{cc}
0 & -2 \\
2 & 0
\end{array}\right). \label{eq:symp1}
\end{equation}
As a consequence, by comparing (\ref{eq:symp1}) and
(\ref{eqn-symp-can}) we see that the scaling factor $\kappa$ is given by
$\kappa=\frac{1}{2}$, that is, $\symp=\frac{1}{2}\omega$, and we
conclude that the K\"ahler metric associated with the symplectic
structure appearing in the canonical equations of motion is given
by $G=\half g$.

\end{document}